\documentstyle[amssymb,preprint,prb,aps]{revtex}
\tightenlines

\begin{document}
\title{Nonstoichiometry, Defects and Transport Properties in MgB$_{2}$ }
\author{Y. Y. Xue, R. L. Meng, B. Lorenz, J. K. Meen, Y. Y. Sun and C. W. Chu$^{%
\text{*}}$}
\address{TCSUH, University of Houston, Houston, TX 77204-5932 ($^{\text{*}}$also at\\
LBNL, 1 Cyclotron Road, Berkeley, CA 94720)}
\date{\today}
\maketitle

\begin{abstract}
The local composition Mg:B of MgB$_{2}$ powder was systematically changed
through annealing. Correlations were observed between the Mg loss and the
lattice parameters {\it a/c} and the microstrain of MgB$_{2}$. Direct
wavelength-dispersive-spectrum and XRD measurements on ceramic samples with
different residual resistivity ratios suggested that the transport property
of MgB$_{2}$ might be affected by this Mg-deficiency, but oxygen-related
defects may play a more important role.
\end{abstract}

One unsolved issue in the transport property of the newly discovered
superconductive MgB$_{2}$ is its apparent sample-dependency. The reported
resistivity at room-temperature, $\rho $(297 K), ranges from 9.6 to 100 $\mu
\Omega $cm, and that at 40 K, $\rho $(40 K), from 0.38 to 21 $\mu \Omega $cm.
\cite{can,jung} Both the $\rho $(297 K) and the residual resistivity ratio,
RRR = $\rho $(297 K)/$\rho $(40 K), vary by a factor of ten among ceramic
samples purported to be high-quality. In a closely related way, the reported
magnetoresistance, $\Delta \rho /\rho $ = [$\rho $(40 K, 5 T)-$\rho $(40 K,
0 T)]/$\rho $(40 K, 0 T), of ceramic MgB$_{2}$ samples varies from 
\mbox{$<$}%
1\% to 60\% ({\it i.e.} more than 100 times),\cite{bud} and the temperature
dependency of the resistivity varies from T$^{2}$ to T$^{3}$.\cite{kang}
While the $\rho $(297 K) of a ceramic sample may depend on its porosity and
grain coupling, the large variations in RRR and $\Delta \rho /\rho $ may
require dense lattice defects. Resistances of grain-boundaries typically
will only add a constant in $\rho $, which should lead to a linear
correlation of $\Delta \rho /\rho $ $\propto $ 1/RRR instead of the $\Delta
\rho /\rho $ $\propto $ (1/RRR)$^{2}$ observed.\cite{chen} The sample
dependency has been previously attributed to either the pressure or the
thermal history used during the synthesis.\cite{kang,cun} However, the fact
that similar low RRR and negligible magnetoresistance have been observed in
samples synthesized both with and without high-pressures challenges the
interpretation. The mechanism responsible for the differences needs to be
explored.

It is interesting to note that the local composition,{\it \ i.e.} both the
atomic ratio B:Mg and the possible oxygen contamination, is still a topic of
debate. Cooper {\it et al.} proposed in 1970 that the boron in borides AB$%
_{2}$ might be nonstoichiometric,\cite{coo} and Zhao {\it et al.} reported
that the superconducting temperature, T$_{C}$, of MgB$_{2}$ changes
significantly with different starting compositions of Mg:B.\cite{zhao} Most
MgB$_{2}$ samples, it should be pointed out, were synthesized with a
stoichiometric starting composition. Noticeable amounts of MgO and/or Mg are
either frozen inside the samples or deposited on the walls of the sample
containers. MgB$_{2}$ samples, therefore, could be non stoichiometric. The
facts that Au foil, which is typically used in high-pressure synthesis, can
easily absorb Mg, and the samples with large RRR were synthesized using
excess Mg,\cite{can} may further enhance the suspicion. On the other hand,
the EELS data of Zhu{\it \ et al.} shown that the local Mg:B is
stoichiometric in their samples with a variation less than $\pm $ 2\%.\cite
{zhu} The Mg-stoichiometry and its effect on transport property, therefore,
may still be open questions.

Oxygen contamination has also been widely discussed, in particular in terms
of the T$_{C}$ degrading of MgB$_{2}$ films.\cite{zhai} However, the T$_{C}$
of a ceramic with a grain size of 1 $\mu $m or larger typically will not be
affected unless the contamination causes lattice defects and/or microstrain.
Zhu{\it \ et al.} have shown that MgO can exist inside MgB$_{2}$ grains as
stacking faults and create microstrains.\cite{zhu} The effects of these
faults, consequentially, will depend on their morphology (therefore, oxygen
concentration, diffusion and synthesis temperatures). It should be
cautioned, that the Mg powder/chips used may also function as an oxygen
getter, and the oxygen related defects may also be affected by the starting
composition.

To explore the topic, we measured X-ray diffraction (XRD), composition and
resistivity of several MgB$_{2}$ ceramic samples. A systematic correlation
among RRR, the lattice distortion (microstrain) and intragrain
oxygen-concentration was observed. The local ratio of Mg:B of two ceramic MgB%
$_{2}$ samples with rather different RRR, however, is almost the same. These
suggest that the transport property of MgB$_{2}$ is dominated by the oxygen
related defects. It should be pointed out, however, a $\leq $ 5\% Mg off
stoichiometry was observed in MgB$_{2}$ powders annealed at low
temperatures. An accompanied systematic change of the lattice
parameters/microstrain further suggested that this off stoichiometry may
also create lattice defects, although the fast diffusion of Mg/B may make
such defects disappear and play only minor roles in typical MgB$_{2}$
ceramic.

Ceramic MgB$_{2}$ samples were prepared using the solid-state reaction
method.\cite{bud} A mixture of small Mg chips (99.8\% pure) and B powder
(99.7\% pure) were sealed inside a Ta tube under an Ar atmosphere. The
sealed Ta ampoule was then enclosed in a quartz tube. The assembly was
heated slowly up to 950~$^{\text{o}}$C followed by rapid quenching. Various
starting Mg:B ratios were used to obtain samples with different RRR's.
Unfortunately, the impurity phases in the samples so prepared make a
quantitative Mg stoichiometry-control difficult. Several Mg-deficient
samples, therefore, were made by vacuum annealing commercial MgB$_{2}$
powder (Alfa, 98\% pure). The structure was determined by X-ray powder
diffraction (XRD) using a Rigaku DMAX-IIIB diffractometer with a Cu target.
The refinement was done using the Reitan-94 program.\cite{izu} The
resistivity was measured through the standard 4-probe method using a R-700 
{\it ac} bridge. Analysis of the magnesium boride grains was done by
electron microprobe analysis (EMPA) at an accelerating voltage of 15 kV and
30 nA beam current using a JEOL JXA 8600 electron microprobe with attached
Wavelength Dispersive Spectrometers (WDS).

To answer the question of whether MgB$_{2}$ can be microscopically
non-stoichiometric, commercial Alfa MgB$_{2}$ powder was annealed in
pre-vacuumed quartz tubes at various temperatures between 700 and 900 $^{%
\text{o}}$C followed by rapid quenching. The weight loss during the
annealing was measured. X-ray powder diffraction was then performed and the
data refined as a mixture of MgB$_{2}$ (space group I-191) and MgO (I-225)
with only the line-profile parameters, the phase composition MgO/MgB$_{2}$,
the lattice constants of MgB$_{2}$, and the isotropic thermal factor
adjustable. The lines of impurity phases other than MgO were not used during
the refinement. A typical data set is shown in Fig. 1. Typically, the
R-pattern, $R_{p}$, the goodness-of-fit, $S$, and the R-structure factor of
MgB$_{2}$, $R_{F}$, are 4, 1.4, and 2, respectively, an indication of
reasonable agreement.\cite{yo} It was noted, however, that the spread of the
refined parameters was larger than the mathematical uncertainty $\sigma $
given by the program. The lattice constants, for example, can spread over $%
\pm $ 0.0005 \AA\ from one sample to another with the same Mg deficiency
while the $\sigma $ is only $\pm $ 1 to 2$\cdot 10^{-4}$ \AA . To estimate
the experimental resolution, therefore, the samples with the same weight
loss were prepared and measured repeatedly. The data spread was used as the
experimental resolution in following figures. During the annealing, the
fraction of MgO seemed to increase from 8\% of the Alfa powder to $\approx $
14\% of the annealed samples. The weight loss and the ratio MgO/MgB$_{2}$,
therefore, were combined to calculate the change of the Mg composition in MgB%
$_{2}$, {\it i.e.} the 1-x in Mg$_{1-x-xo}$B$_{2}$, assuming that the Alfa
powder is Mg$_{1-xo}$B$_{2}$. Other impurity phases, such as MgB$_{4}$,
appear when 1-x 
\mbox{$<$}%
0.86 and this estimation can no longer be used. It should be noted that it
might also be possible that the Mg deficiency caused by the formation of MgO
might be spatially very inhomogeneous, and the 1-x assumed could have a
systematic shift as large as 0.06.

A noticeable trend was observed in a decrease of the lattice parameter {\it c%
}, and an increase of {\it a} with increasing of 1-x (Fig. 2). To verify
that this change is not due to some unknown effect of the heat treatment,
metal Mg chips were added to one of the quartz tubes to prevent Mg-loss
during the annealing. The weight loss and the change in the MgO fraction
observed were indeed the same as those of the initial Alfa powder. The {\it %
a }and {\it c} of this Mg-added sample (the triangles in Fig. 2) are
identical to those of the Alfa powder, demonstrating that Mg-loss during the
anneal is the main cause of the change in lattice parameters. This
systematic change of {\it a/c} will not be possible without a noticeable
off-stoichiometry of Mg, which is different from the almost ideal
composition observed by Zhu {\it et al,} \cite{zhu} although the missing or
intercalation of boron/magnesium planes observed there show that an
off-stoichiometry should be possible in principle. We attribute it to the
differences in sample-preparation, {\it e.g}. the lower annealing
temperature used here may freeze the Mg- and B-related defects. It should be
noted that no significant variation in {\it a}/{\it c} was observed with 1-x 
\mbox{$<$}%
0.85, suggesting a phase-stability boundary around Mg:B $\approx $ 0.85:2.
Variation of {\it a}/{\it c} was reported previously with a much broader
Mg:B range (0.5:2 to 2:2).\cite{zhao} Significant impurity phases, {\it e.g.}
MgB$_{4}$, in the samples may cause the difference based on the reported XRD.

A more subtle but important change associated with the Mg-loss is the
broadening of the XRD lines. The line width significantly increases with the
loss of Mg (Fig. 3), although it is already much larger than our
instrumental resolutions in the initial Alfa powder. It should be noted that
the true change in the line profile would be much more prompt if there were
no interference between the Cu K$_{\alpha 1}$ and K$_{\alpha 2}$ lines. The
significant contribution from the K$_{\alpha 2}$ line, unfortunately, makes
a detailed analysis of the anisotropic strain impractical. We, therefore,
have to depend on the Rietveld refinement. An isotropic broadening was
assumed, {\it i.e.} the line profile was fitted with the pseudo-Voigt
function of the scattering angle 2$\theta $.\cite{yo} Mathematically, the
function is a convolution of a Lorentz function with a Gauss function. To
deduce the instability of the nonlinear regression, the variance $\sigma ^{2}
$ of the Gauss function was taken as {\it U}tan$^{2}\theta $+{\it V}tan$%
\theta $+{\it W, i.e. }ignoring the Scherrer term {\it P}sec$^{2}\theta $
since it is rather small anyway. The parameters V and W, which are not
expected to be sample-dependent, were further fixed to the values obtained
in a calibration run with well-crystallized Al$_{2}$O$_{3}$ powder.
Similarly, the full-width-at-half-maximum (FWHM) of the Lorentz function H$%
_{kL}$ was assumed to be {\it X}$\cdot $sec$\theta $+{\it Y}$\cdot $tan$%
\theta $, {\it i.e.} ignoring the anisotropic broadening terms {\it X}$_{e}$
and {\it Y}$_{e}$.\cite{yo} Previously, the parameters {\it U }and{\it \ Y,
i.e. }the components varying as tan$\theta $,{\it \ }were used to
characterize the microstrain, {\it i.e.} either a modulation of local
lattice parameters or their domain-to-domain variation. The sec$\theta $
component {\it X}, on the other hand, was generally used to describe the
size of the structural domains.

The {\it U} and {\it Y} of the annealed powders are shown in Fig. 4. While $U
$ seems to increase with x linearly over the whole range probed, $Y$ seems
to be noticeably affected only beyond a threshold Mg loss of x $\approx $
0.1. The line profile seems to be dominated by the Gauss components in the
slightly Mg-deficient samples, but by the Lorentz components in the heavily
Mg-deficient ones. The value of {\it X,} on the other hand,{\it \ }shows no
systematic change with x. The widening of the XRD lines, therefore, is
mainly related to the local variation of the lattice parameters. We further
attribute it to the non-stoichiometry of MgB$_{2}$ based on the systematic
changes of the {\it a} and {\it c} with the x observed. To qualitatively
estimate this microstrain, the overall FWHM of the pseudo-Voigt functions
with the $U$ and $Y$ equal to the observed values was calculated as H$_{k}$,
and that with $U$ = $Y$ = 0 as H$_{k0}$. The contribution of microstrain was
then taken as $\Delta \theta $=$\sqrt{\text{H}_{\text{k}}^{\text{2}}\text{-H}%
_{\text{k0}}^{\text{2}}}$. The weighted relative variation of the {\it a}
and {\it c} (the microstrain), therefore, should be of the order of $%
\varepsilon =\Delta \theta /$tan$\theta $. The microstrain so-calculated is
almost $\theta $-independent between $\theta $ = 10$^{\text{o}}$ and 40$^{%
\text{o}}$ in a given sample, as expected. A systematic increase of $%
\varepsilon $ with the loss of Mg was observed (Fig. 5). It seems that the
Mg-loss is extremely non-uniform, and the resulting local defects could
serve as scattering centers for the carriers.

Despite their differences in composition and lattice parameters, however,
the superconductivity measured on these samples were almost the same (Fig.
6). The onset of the zero-field-cooled magnetization, M$_{ZFC}$, is 39 K and
the transition width (10\% to 80\% of M$_{ZFC}$ at 5 K) differs by less than
0.5 K from sample to sample. This may be understood in terms of percolation
and the fact that MgB$_{2}$, as a S-wave BCS superconductor, should not be
sensitive to defects. The observation that Mg nonstoichiometry alone will
not cause large T$_{C}$ change is also supported by our data on several
samples synthesized with starting composition Mg:B from 0.5:2 to 2.4:2.\cite
{meng} It is interesting to note that the field-cooled magnetizations, M$%
_{FC}$, were also sample-independent. The similar vortex pinning strength
suggests that the length scale between defects might be comparable or
smaller than the coherent length ($\approx $ 60 \AA\ for MgB$_{2}$).

Unfortunately, no transport measurement can be made in these annealed
samples, which were made in the powder form to ensure the uniform loss of
Mg. Therefore, additional ceramic samples were purposely made with different
starting compositions. These samples show different RRR. The $\rho $(T) of
two typical samples with RRR = 2.5 (Sample A) and 7 (Sample B),
respectively, are shown in Fig. 7. Sample B was synthesized with extra
Mg-chips and a start composition of Mg:B=1.25:2. It is expected, therefore,
that Sample B should be relatively Mg-rich and O-free.

The {\it c (a)} is 3.525 and 3.519 \AA\ (3.084 and 3.085 \AA ) observed for
Sample A and B, respectively. While the shorter {\it c} of Sample B{\it \ }%
seems to be in qualitative agreement with the assumption that it is Mg-rich,
the similar {\it a} observed suggests that other defects may also play a
role. In fact, the estimated 1-x of Sample B would be 1.2 based on its {\it c%
}, but 1.03 based on the {\it a} using Fig{\it . }2 as the calibration curve%
{\it . }To further explore the topic, {\it a/c }of several ceramic samples
with RRR from 2.3 to 7 were measured (Fig. 8). When the {\it c} appears to
decrease systematically with RRR, no clear correlation exists between the 
{\it a} and RRR. Defects other than Mg-deficiency seem to play a role, based
on a comparison between Figs. 2 and 8.

To verify the Mg stoichiometry in Samples A and B, the samples were
subjected to EMPA analysis for Mg, B, Si, and O, together with a
cool-pressed Alfa MgB$_{2}$ powder. The observed ratio B:Mg $\approx $
2.01:1 for the Alfa powder sample has a standard deviation of $\pm $0.04 ($%
\approx $2\%) over 12 positions measured. Similarly, the ratio is 1.99$\pm $%
0.03:1 for Sample A, but 2.06$\pm $0.17:1 for Sample B. The greater
variability of the composition of Sample B is plausibly the result of its
much smaller grain size (%
\mbox{$<$}%
2 $\mu $m in diameter). The regions in the grain boundaries introduces an
additional complicating factor. It is interesting to note, however, that the
point-to-point variation in Sample A and the Alfa powder still exceeds the
precision of our analysis ($\approx $ 0.4\%), although their grains are much
larger. This may be an indication of a real composition variability, and in
agreement with the nonstoichiometry discussed above. In such a case, the
data after the average should be much closer to the true value. The almost
identical average compositions of 1-x $\approx $ 1.01 and 0.97 in Samples A
and B, respectively, show that the nonstoichiometry may not be the
dominating factor for the different RRR observed in ceramic samples. It is
interesting to note that the Mg off-stoichiometry of these as-synthesized MgB%
$_{2}$ ceramic samples is rather small, which is in agreement with the EELS
data of Zhu {\it et al}.\cite{zhu}

However, a clear trend was observed between RRR and the microstrain $%
\varepsilon $. The raw XRD data already show that the line width of Sample B
is significantly narrower (Fig. 3). The refinement further demonstrate that
most of the narrowing comes from the decrease of $U$ and $Y$ (the
microstrain terms). In fact, the combined $\varepsilon $ is 0.004, 0.01, and
0.008 for Samples A and B and the Alfa powder, respectively. To verify the
assumption, the $\varepsilon $ of several ceramic samples with RRR from 2.2
to 7 was also measured (Fig 9). The $\varepsilon $ decreases with RRR, and a
strain 
\mbox{$<$}%
0.2\% seems to be needed for RRR $\approx $ 20. The residual resistance,
therefore, seems to be related to a lattice distortion other than Mg
nonstoichiometry.

Indeed, a correlation was observed between the RRR and local oxygen
composition in several ceramic MgB$_{2}$ samples. The atomic ratio O/Mg, for
example, was 0.15 $\pm $ 0.03 and 0.08 $\pm $ 0.03 for Samples A and B,
respectively. It should be pointed out that the extra oxygen observed in
Sample A is unlikely to come from separated MgO grains based on the observed
grain size (1-5 $\mu $m) and the estimated space resolution (see appendix).
It is also difficult to attribute the contamination to surface oxide layers.
Sample B, for example, has much smaller grain size and should have severe
oxygen contamination if the oxygen is mainly from grain surfaces. We,
therefore, suggest that oxgen may exist inside MgB$_{2}$ grains as defects,
which is in agreement with the MgO/MgB$_{2}$ stacking faults observed by Zhu 
{\it et al}.\cite{zhu} To verify that, the ratio O:Mg in other five samples
were measured (Fig. 10). The trend is clear, although the data scattering is
large.

In summary, our data suggest that MgB$_{2}$ can be nonstoichiometric up to
5-10\%. Although this nonstoichiometry may affect the transport properties,
oxygen related defects seem to play dominating roles in MgB$_{2}$.

ACKNOWLEDGMENT This work is supported in part by NSF Grant, the T. L. L.
Temple Foundation, the John and Rebecca Moores Endowment and the State of
Texas through TCSUH, and in LBNL by DOE.

{\large APPENDIX}

Analysis of the magnesium boride grains was done by electron microprobe
analysis (EMPA) using an accelerating voltage of 15 kV and 30 nA beam
current. Samples were set in epoxy and polished to 1 $\mu $m relief. Primary
standards used were MgO for Mg and O, but pure boron for B. Peaks were
counted for 100 s and two backgrounds for 50 s each for each element. A 20 $%
\mu $m spot size was used for standards but a focused beam was used on the
sample grains. Calculations based on electron-target atom interactions
suggest that 
\mbox{$>$}%
99\% of characteristic x-rays for samples are derived from a region with a $%
\approx $ 2.5 $\mu $m diameter. Characteristic x-rays derived from boron
atoms with an energy of $\approx $183 eV are derived from the top 1.4 $\mu $%
m of the sample. Those of Mg (1253 eV) are from the top 1.7 $\mu $m. Oxygen
x-ray is from the top 2.0 $\mu $m. Thus, the volumes sampled by the x-rays
are similar for each element and are $\approx $10 $\mu $m$^{3}$.

Characteristic x-rays of each element were separated from other photons
using wavelength-dispersive spectrometers employing a crystal of thallium
acid phthalate (Mg) and synthetic layered dispersive elements (B, O). The
energy of the Mg K$\alpha $ x-rays are identical within reproducibility for
standards and samples. The characteristic x-rays of both B and O do,
however, indicate significant changes in energy between samples and
standards and even between different orientations of the magnesium boride
crystalline samples. This reflects differences in the energy levels occupied
by the 2{\it p} electrons in these atoms. The spectrometer position required
to maximize the characteristic B K$\alpha $ x-ray was, therefore, determined
for each sample before each analysis. Careful determination of the FWHM of
the x-ray peak indicated that the shape of the peak (as opposed to the mean
energy) did not vary significantly from the pure B standard to the samples
nor between the different samples.

The raw x-ray intensities were corrected using the on-line Geller iterative
procedure. The resultant corrections were very large for B, in particular.
Thus, the corrected fraction of boron in a sample was computed to be
approximately ten times the fraction of boron estimated by comparing raw
intensities of x-rays on sample and standard. The corrected fraction of Mg
is typically 0.5 the raw estimate. These corrections reflect the large
differences in inter-element correction factors between standards and
samples. Commercial MgB$_{2}$ was therefore used as a secondary standard for
all analyses. The secondary standard was carbon-coated at the same time as
the samples in each case. Routine techniques allow control of the thickness
of the carbon coat to within $\approx $10\% of the nominal 200 nm. The
marked absorption of characteristic B x-rays by carbon warrants an even
higher level of control and this was monitored by determining the intensity
of boron x-rays emitted from the secondary standard.

\begin{figure}[t]
\caption{The XRD pattern of the initial MgB$_{2}$ powder (98\%, Alfa). $%
\cdot $: data; top solid line: Rietveld refinement; bottom solid line:
residual; \#: lines of MgO; *: lines of unknown impurity.}
\label{fig1}
\end{figure}

\begin{figure}[t]
\caption{The lattice parameters {\it a} (open symbols) and {\it c} (filled
symbols). $\bigtriangledown ,$ $\blacktriangledown $: initial Alfa powder; $%
\bigcirc $, $\bullet $: annealed in vacuumed tube; $\Box ,\blacksquare $:
annealed together with Mg-chips.}
\label{fig2}
\end{figure}

\begin{figure}[t]
\caption{Line-profiles of the 002 line of MgB$_{2}$ at 51.8$^{\text{o}}$.
solid line: the initial Alfa powder; dashed line: an annealed powder; dotted
line: a sample with a resistivity ratio RRR = 5; dotted-dashed line: the 024
line of a Al$_{2}$O$_{3}$ sample at 52.3$^{\text{o}}$ to show the
instrumental resolution. The lines have been parallel shifted and normalized
to emphasize the line width. }
\label{fig3}
\end{figure}

\begin{figure}[t]
\caption{The profile parameters $U$ (open symbols) and $Y$ (filled symbols). 
$\bigtriangledown ,$ $\blacktriangledown $: initial Alfa powder; $\bigcirc ,$
$\bullet $: annealed in tube; $\Box $, $\blacksquare $: annealed together
with Mg-chips. }
\label{fig4}
\end{figure}

\begin{figure}[t]
\caption{The estimated isotropic microstrain $\protect\varepsilon $ vs.
Mg-loss of 1-x. $\bigtriangledown $: initial Alfa powder; $\bigcirc $:
annealed in tube; $\Box $: annealed together with Mg-chips. }
\label{fig5}
\end{figure}
\begin{figure}[t]
\caption{M$_{ZFC}$ (open symbols) and M$_{FC}$ (filled symbols) of two
sample with 1-x = 0 ($\bigcirc ,$ $\bullet $) and 0.85 ($\bigtriangleup $, $%
\blacktriangle $), respectively. }
\label{fig6}
\end{figure}

\begin{figure}[t]
\caption{$\protect\rho $(T) of two typical samples. +: data; solid lines:
a+bT$^{n}$ fits. }
\label{fig7}
\end{figure}

\begin{figure}[t]
\caption{The lattice parameters {\it a} ($\bullet $) and {\it c} ($%
\blacktriangledown $) as functions of RRR. }
\label{fig8}
\end{figure}
\begin{figure}[t]
\caption{The strain $\protect\varepsilon $ {\it vs}. RRR}
\label{fig9}
\end{figure}
\begin{figure}[t]
\caption{Oxygen concentration O/Mg {\it vs}. RRR}
\label{fig10}
\end{figure}


\begin{references}
\bibitem{can}  P. C. Canfield, D. K. Finnemore, S. L. Bud'ko, J. E.
Ostenson, G. Lapertot, C. E. Cunningham and C. Petrovic, cond-mat/0102289

\bibitem{jung}  C. U. Jung, M. S. Park, W. N. Kang, M. S. Kim, S. Y. Lee and
S. I. Lee, cond-mat/0102215.

\bibitem{bud}  S. L. Bud'ko, C. Petrovic, G. Lapertot, C. E. Cunningham, P.
C. Canfield, M. H. Jung, A. H. Lacerda, cond-mat/0102413.

\bibitem{kang}  W. N. Kang, C. U. Jung, K. H. P. Kim, M. S. Park, S. Y. Lee,
H. J. Kim, E. M. Choi, K. H. Kim, M. S. Kim and S. I. Lee, cond-mat/0102313.

\bibitem{chen}  X. H. Chen, Y. S. Wang, Y. Y. Xue, R. L. Meng, Y. Q. Wang
and C. W. Chu, unpublished

\bibitem{cun}  C. E. Cunningham, C. Petrovic, G. Lapertot, S. L. Bud'ko, F.
Laabs, W. Straszheim, D. K. Finnemore and P. C. Canfield, cond-mat/0103390.

\bibitem{coo}  A. S. Cooper, E. Corenzwit, L. D. Longinotti, B. T. Matthias
and W. H. Zachariasen, Proc. Natl. Acad. Sci. 67, 313 (1970).

\bibitem{zhao}  Y. G. Zhao, X. P. Zhang, P. T. Qiao, H. T. Zhang, S. L. Jia,
B. S. Cao M. H. Zhu, Z. H. Han, X. L. Wang and B. L. Gu, cond-mat/0105053.

\bibitem{zhu}  Y. Zhu, L. Wu, V. Volkov, Q. Li, G. Gu, A. R. Moodenbaugh, M.
Malac, M. Suenaga and J. Tranquada, cond-mat/0105311.

\bibitem{zhai}  H. Y. Zhai, H. M. Christen, L. Zhang, M. Parathaman, C.
Cantoni, B. C. Sales, P. H. Fleming, D. K. Christen and D. M. Lowndes,
cond-mat/0103618.

\bibitem{izu}  F. Izumi, Rigaku J. 6, 10 (1989).

\bibitem{yo}  For example, ''The Rietveld Method'', edited by R. A. Young
(New York, Oxford University Press, 1993).

\bibitem{note}  The value of {\it P} is rather small when set as free
parameters in some test runs, and not expected to affect our conclusion
significantly.

\bibitem{meng}  R. L. Meng {\it et al} unpublished.
\end{references}
\end{document}